\documentclass[
preprintnumbers,
preprint,
a4paper,
showpacs,
amsmath,
amssymb,
byrevtex
]{revtex4}
\usepackage{graphicx}
\begin{document}
\title{Black hole formation in the grazing collision \\
  of high-energy particles }
\author{Hirotaka Yoshino}
\email{hyoshino@allegro.phys.nagoya-u.ac.jp}
\author{Yasusada Nambu}
\email{nambu@allegro.phys.nagoya-u.ac.jp}
\affiliation{Department of Physics, Graduate School of Science, Nagoya
University, Chikusa, Nagoya 464-8602, Japan}
\preprint{DPNU-02-26}
\date{\today}
\begin{abstract}
  We numerically investigate the formation of $D$-dimensional black
  holes in high-energy particle collision with the impact parameter and
  evaluate the total cross section of the black hole production.  We find
  that the formation of an apparent horizon occurs when the distance
  between the colliding particles is less than $1.5$ times the
  effective gravitational radius of each particles.  Our numerical
  result indicates that although both the one-dimensional hoop and the
  $(D-3)$-dimensional volume corresponding to the typical scale of the
  system give a fairly good condition for the horizon formation in the
  higher-dimensional gravity, the $(D-3)$-dimensional volume provide a
  better condition to judge the existence of the horizon.
\end{abstract}
\maketitle
\section{Introduction}

The brane world scenario is paid much attention in the context of the
unified theory of elementary particles. This scenario regards our
space as the 3-brane with large extra dimensions on which gauge
particles and interactions are confined. In this scenario, the Planck
energy can be at the $O(\text{TeV})$ scale~\cite{TSS}. One of the
consequences of lowering the Planck scale is that the properties of a
small black hole whose radius is smaller than the size of the extra
dimensions are substantially altered; the black holes is
well-described as a $D$-dimensional black hole centered on the brane,
but extending out into the $(D-4)$-dimensional extra space and the
radius of such a black hole is $\sim 10^{32}$ times larger than that
of the usual black hole with the same mass. Hence it becomes much
easier to produce black holes using a future planned accelerator such
as the CERN Large Hadron Collider and this possibility has been
discussed by many authors~\cite{BHUA}.

The possibility of producing black holes by the collision of particles
can be estimated by using the hoop conjecture~\cite{Th72}, which
states that an apparent horizon forms when and only when the mass $M$
of the system gets compacted into a region whose circumference $C$
satisfies
\begin{equation}
  H_D\equiv \frac{C}{2\pi r_h(M)}\lesssim 1,
  \label{eq:hc}
\end{equation} 
where $r_h(M)$ is the Schwarzschild horizon radius for the mass $M$.
If we assume that the inequality \eqref{eq:hc} give the condition for
black hole formation in the higher-dimensional gravity, we can
evaluate the impact parameter $b$ of the colliding particles which
leads to black hole production.  The circumference that surrounds two
particles at the instant of collision is $C\sim 2b$. By setting
$2b/2\pi r_h(2\mu)\sim 1$ where $2\mu$ is the center of mass energy of
the system, the maximal impact parameter $b_{\text{max}}$ that leads
to the black hole formation becomes $b_{\text{max}}\sim r_h(2\mu)$. By
introducing a numerical factor $F(D)$ close to unity, the total cross
section for black hole production is written as the following form:
\begin{equation}
  \sigma_{\text{b.h.production}}=F(D)\pi r_h^2(2\mu).
  \label{eq:crosssection}
\end{equation}
Obtaining $F(D)$ is necessary to improve experimental predictions of
the black hole production in collider physics and observations of
ultra-high energy cosmic rays.

In our previous paper~\cite{YN02}, we investigated the black hole
formation in high-energy head-on collisions of particles in the
$D$-dimensional gravity and found that $H_D$ becomes a parameter
to judge the existence of the apparent horizon for all $D$. 
Although this implies that the hoop
conjecture holds for this system, we do not know the condition for the
horizon formation in the higher-dimensional spacetime.  Recently, Ida
and Nakao~\cite{IN02} investigated momentarily static initial data in
the five-dimensional spacetime and found that a spindle distribution of
matter can lead to the black hole formation even if the hoop $C$ is
arbitrarily larger than $2\pi r_h(M)$. This provides a counter example
for ``only when'' part of the hoop conjecture. They showed that the
isoperimetric inequality $(\text{characteristic area})\lesssim 4\pi
r_h^2(M)$ is satisfied on the horizon and conjectured that for a
characteristic $(D-3)$-dimensional volume $V_{D-3}$ of the system, the
inequality
\begin{equation}
  \label{eq:vc}
  \frac{V_{D-3}}{G_D M}\lesssim 1
\end{equation}
would become a condition for the horizon formation in the
$D$-dimensional spacetime with the gravitational constant $G_D$.  This
``$(D-3)$-volume conjecture''(volume conjecture) is one candidate that
provides a better condition for the horizon formation in the
higher-dimensional spacetime than the hoop conjecture and it is worth
investigating.

In this paper, we consider the grazing collisions of particles in the
$D$-dimensional Einstein gravity and investigate the formation of
apparent horizons.  Eardley and Giddings~\cite{EG02} developed a
method of finding apparent horizons for this system.  The problem was
reduced to a boundary-value problem for a Poisson's equation in a
$(D-2)$-dimensional flat space and they solved it analytically for
$D=4$ case. We solve this problem numerically for $D>4$ and obtain the
maximal impact parameter $b_{\text{max}}$ for black hole formation and
the factor $F(D)$ in Eq. \eqref{eq:crosssection}.  Then using these
solutions, we discuss the condition of horizon formation for this
system from the view point of the hoop conjecture and the volume
conjecture.

This paper is organized as follows. In Section II, we briefly review
the method of finding apparent horizons in the system of the grazing
collision of particles.  In Section III, we present our numerical
results and discuss the two conjectures for this system. Section IV is
devoted to summary and discussion.

\section{ Apparent horizons in the grazing collision of high-energy particles}

To simplify the situation, we consider black holes with horizon radius
smaller than the size of extra dimensions. This enables us to ignore
the effect of the brane tension and the geometry of the extra
dimensions. The metric with a high-energy point particle is obtained
by infinitely boosting a Schwarzschild black hole with fixed total
energy $\mu$. The resulting system becomes a massless point particle
accompanied by a plane-fronted gravitational shock wave which is the
Lorentz-contracted longitudinal gravitational field of the particle.
Combining two shock waves, we can set up the collision of high-energy
two particles moving in $\pm z$ direction. This system was originally
developed by D'Eath and Payne~\cite{DEP92} and recently analyzed
in~\cite{EG02,YN02}. The metric of this system outside the future
light cone of the colliding shocks is given as
\begin{align}
  ds^2&=-dudv
  +\left(H^{(+)}_{ik}H^{(+)}_{jk}+H^{(-)}_{ik}H^{(-)}_{jk}
    -\delta_{ij}\right)dx^idx^j, \notag \\
  & H^{(+)}_{ij}=\delta_{ij}+\frac{u}{2}\Theta(u)
  \nabla_i\nabla_j\Phi
  (\boldsymbol{x}-\boldsymbol{x}_+),\\
  & H^{(-)}_{ij}=\delta_{ij}+\frac{v}{2}\Theta(v)
  \nabla_i\nabla_j\Phi
  (\boldsymbol{x}-\boldsymbol{x}_-), \notag
\end{align}
where $u=t-z$, $v=t+z$, $\Theta$ is the Heviside step function and
$\boldsymbol{x}\equiv(x^i)$ is the point in flat $(D-2)$-space
$(x_1,\cdots,x_{D-2})$ that is transverse to the direction of particle
motion. The function $\Phi(\boldsymbol{x})$ depends only on
$r\equiv|\boldsymbol{x}|=\sqrt{{x}_i{x}^i}$ and takes the form
\begin{alignat}{3}
  & \Phi(\boldsymbol{x})=-8G_4\mu\log{r},&\quad\text{for}~D=4 \\
  & \Phi(\boldsymbol{x})=\frac{16\pi\mu
    G_D}{\Omega_{D-3}(D-4)}\frac{1}{{r}^{D-4}},&\quad\text{for}~D>4
\end{alignat}
where $\Omega_{D-3}$ is the volume of a unit $(D-3)$-sphere and
$\boldsymbol{x}_\pm$ denote the points of two particles in
$(D-2)$-dimensional space and we take
\begin{eqnarray}
  \boldsymbol{x}_\pm&=&(\pm b/2,0,\cdots,0).
\end{eqnarray}

The apparent horizon $\mathcal{S}$ is defined as a closed spacelike
$(D-2)$-surface on which the outer null geodesic congruence has zero
convergence.  Eardley and Giddings~\cite{EG02} developed a method of
finding the apparent horizon in the union of the two shock waves, $u=0>v$
and $v=0>u$.  Their method reduces the problem to finding the
$(D-3)$-dimensional closed surface $\mathcal{C}$ and two functions
$\Psi_\pm(\boldsymbol{x})$ on the $(D-2)$-dimensional plane satisfying
\begin{align}
  & \nabla^2\left(\Phi_\pm-\Psi_\pm\right)=0 
  \quad\text{interior to}\quad \mathcal{C}, 
  \label{eq:condition1}\\
  & \Psi_\pm=0\quad\text{on}\quad \mathcal{C},
  \label{eq:condition2}\\
  & \nabla\Psi_+\cdot\nabla\Psi_-=4\quad\text{on}\quad \mathcal{C},
  \label{eq:condition3}
\end{align} 
where $\Phi_\pm \equiv\Phi (\boldsymbol{x}-\boldsymbol{x}_\pm)$.  If
we can find a solution of
Eqs.~\eqref{eq:condition1},\eqref{eq:condition2} and
\eqref{eq:condition3}, a surface with zero expansion is given as the
union of two $(D-2)$-surfaces $v=-\Psi_+(\boldsymbol{x})$ in $u=0>v$
and $u=-\Psi_-(\boldsymbol{x})$ in $v=0>u$. $\mathcal{C}$ is the
intersection of the two $(D-2)$-surfaces at $u=v=0$. For $D=4$,
Eardley and Giddings solved the above problem
analytically~\cite{EG02}.  They used the conformal invariance of
two-dimensional Poisson's equation.  For $D>4$, we cannot use their
method because $(D-2)$-dimensional Poisson's equation does not have
conformal invariance. We must solve the above problem numerically.

In the $(D-2)$-dimensional space, $\mathcal{C}$ and $\Psi_\pm$ have a
symmetry about $x_1$-axis that connects two source points.  Hence
solving the apparent horizon is reduced to a two-dimensional problem.
We use spherical coordinate $(r,\theta)$ to represent the point
$\boldsymbol{x}$ in $(D-2)$-dimensional space, where $r$ is the
distance $|\boldsymbol{x}|$ from the origin and $\theta$ is the angle
between $\boldsymbol{x}$ and $x_1$-axis.  We parameterize the curve
$\mathcal{C}$ by $r=g(\theta)$. The function $g(\theta)$ and
$\Psi_\pm$ have symmetry $g(\theta)=g(\pi-\theta)$ and
$\Psi_+(r,\theta)=\Psi_-(r,\pi-\theta)$.  We calculate $\mathcal{C}$
and $\Psi_\pm$ numerically as follows.  First, we give a trial curve
$\mathcal{C}$ and solve the equation~\eqref{eq:condition1} under the
boundary condition~\eqref{eq:condition2}.  Then, we calculate the
value $\delta=\nabla\Psi_+\cdot\nabla\Psi_--4$ on $\mathcal{C}$, and
determine a modified boundary $\mathcal{C}$ using the value of
$\delta$.  $\mathcal{C}$ and $\Psi_+$ converge to the solution of
Eqs.~\eqref{eq:condition1}~\eqref{eq:condition2}~\eqref{eq:condition3}
by iterating these two steps.  In order to proceed the first step, we
introduce a regular function $h\equiv \Phi_+-\Psi_+$ that obeys
$D$-dimensional Laplace equation
\begin{equation}
  h_{,rr}+\frac{D-3}{r}h_{,r}+\frac{1}{r^2}h_{,\theta\theta}
  +\frac{D-4}{r^2}\cot\theta h_{,\theta}=0.
  \label{eq:Laplace1}
\end{equation}
The boundary condition is
\begin{equation}
  h\big|_{r=g(\theta)}=\Phi_+, \quad
  \frac{\partial h}{\partial\theta}\Big|_{\theta=0}
  =\frac{\partial h}{\partial\theta}\Big|_{\theta=\pi}=0,
\end{equation}
that comes from \eqref{eq:condition2} and the symmetry about
$x_1$-axis.  To simplify the boundary condition, we introduce a new
coordinate $(\tilde{r},\tilde{\theta})$ by
\begin{equation}
  r=g(\tilde{\theta})\tilde{r},\quad
  \theta=\tilde{\theta}.
\end{equation}
In this coordinate, the equation \eqref{eq:Laplace1} becomes
\begin{align}
  &\left(1+\frac{{g^\prime}^2}{g^2}\right)h_{,\tilde{r}\tilde{r}}
  +\frac{1}{\tilde{r}^2}h_{,\tilde{\theta}\tilde{\theta}}
  -\frac{2}{\tilde{r}}\frac{g^\prime}{g}h_{,\tilde{\theta}\tilde{r}}\nonumber\\
  &\qquad+\frac{1}{\tilde{r}}\left(
    D-3-\frac{g^{\prime\prime}}{g}+2\frac{{g^\prime}^2}{g^2}
    -(D-4)\cot\tilde{\theta}\frac{g^\prime}{g}
  \right)
  h_{,\tilde{r}}
  +\frac{D-4}{\tilde{r}^2}\cot\tilde{\theta} h_{,\tilde{\theta}}=0
\label{eq:Laplace2}
\end{align}
and the boundary $\mathcal{C}$ is given by $\tilde{r}=1$.  We solve
\eqref{eq:Laplace2} using the finite differential method with $(50\times
100)$ grids.  There is a coordinate singularity and we cannot write
down the finite difference equation of~\eqref{eq:Laplace2} at
$\tilde{r}=0$. For $\tilde r=0$, we use the Laplace equation in
$(x_1,\rho)$-coordinate
\begin{equation}
  h_{,x_1x_1}+(D-3)h_{,\rho\rho}=0\quad\text{at}\quad \rho\rightarrow 0,
  \label{eq:Laplace3}
\end{equation} 
where $\rho$ is the distance from $x_1$-axis. We relate the value at
the grid points around $\tilde r=0$ using Eq.~\eqref{eq:Laplace3}.

Once $h$ converges, we proceed the second step to determine the
modified boundary $\mathcal{C}$.  We calculate
$\delta(\theta)=\nabla\Psi_1\cdot\nabla\Psi_2-4$, and determine the
modified boundary $\mathcal{C}$ as
\begin{equation}
  r=g(\theta)
  +\epsilon\times \delta(\theta).
\end{equation}
If $\epsilon$ is sufficiently small, $\mathcal{C}$ converges. In our
calculation, we took the value of $\epsilon$ as typically $10^{-3}\sim
10^{-5}$ times the horizon radius of the system.  Smaller $\epsilon$
is required to converge the calculation for the larger impact
parameter $b$.  The maximal impact parameter $b_{\text{max}}$ is
obtained when $\partial g(\theta,b)/\partial b$ becomes $-\infty$.
For $D=10$ and $11$, the numerical instability occurs at the
neighborhood of $b=b_{\text{max}}$. We obtained $b_{\text{max}}$ by
extrapolating $g(\pi/2,b)$ by fitting the numerical data.

To evaluate numerical errors, we compare the numerical solution with
the analytic solution given by Eardley and Giddings for $D=4$ case.
The numerical error is less than $0.02\%$.  For $D\ge 5$, we evaluate
the error by comparing the solution with the result of the calculation
with $(100\times 200)$ grids.  The error is about $0.04\%$ for $D=6$,
$0.07\%$ for $D=8$, $0.25\%$ for $D=10$, and $0.4\%$ for $D=11$.

\section{Numerical results and condition for horizon formation}
\begin{figure}[t]
\centering
{\includegraphics[width=0.4\linewidth]{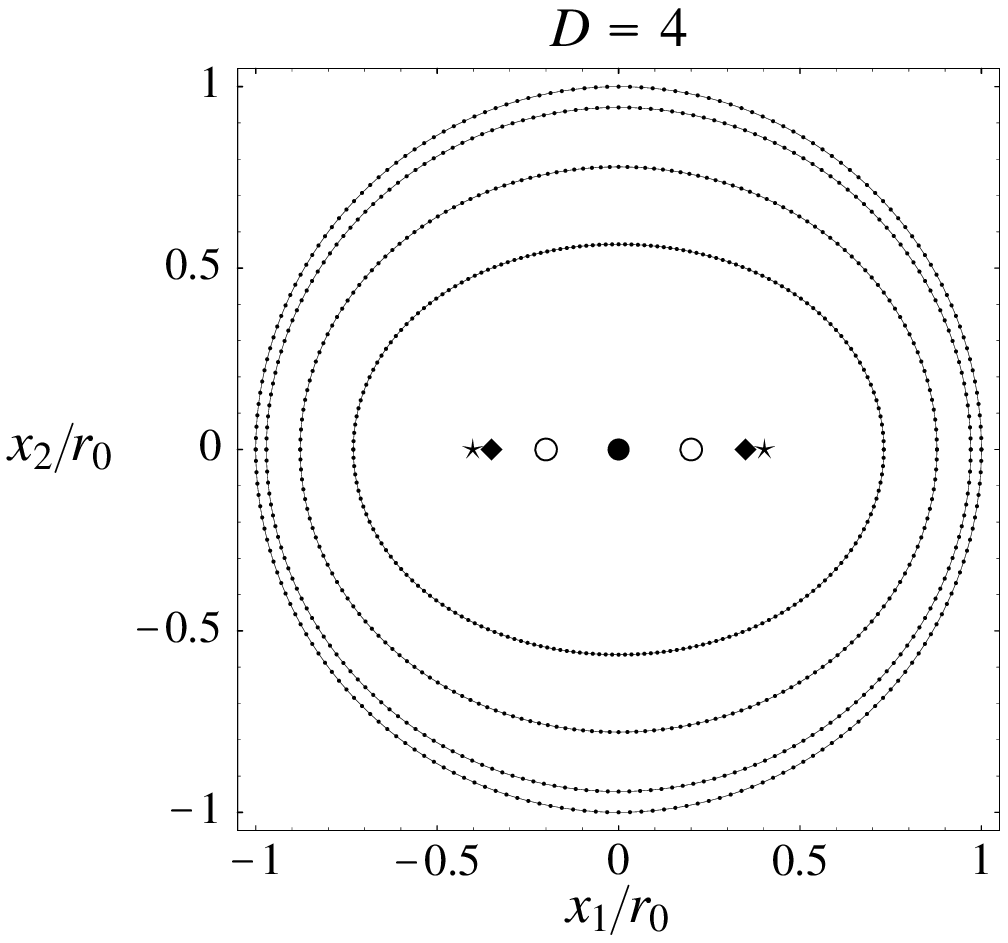}\hspace{10mm}
\includegraphics[width=0.4\linewidth]{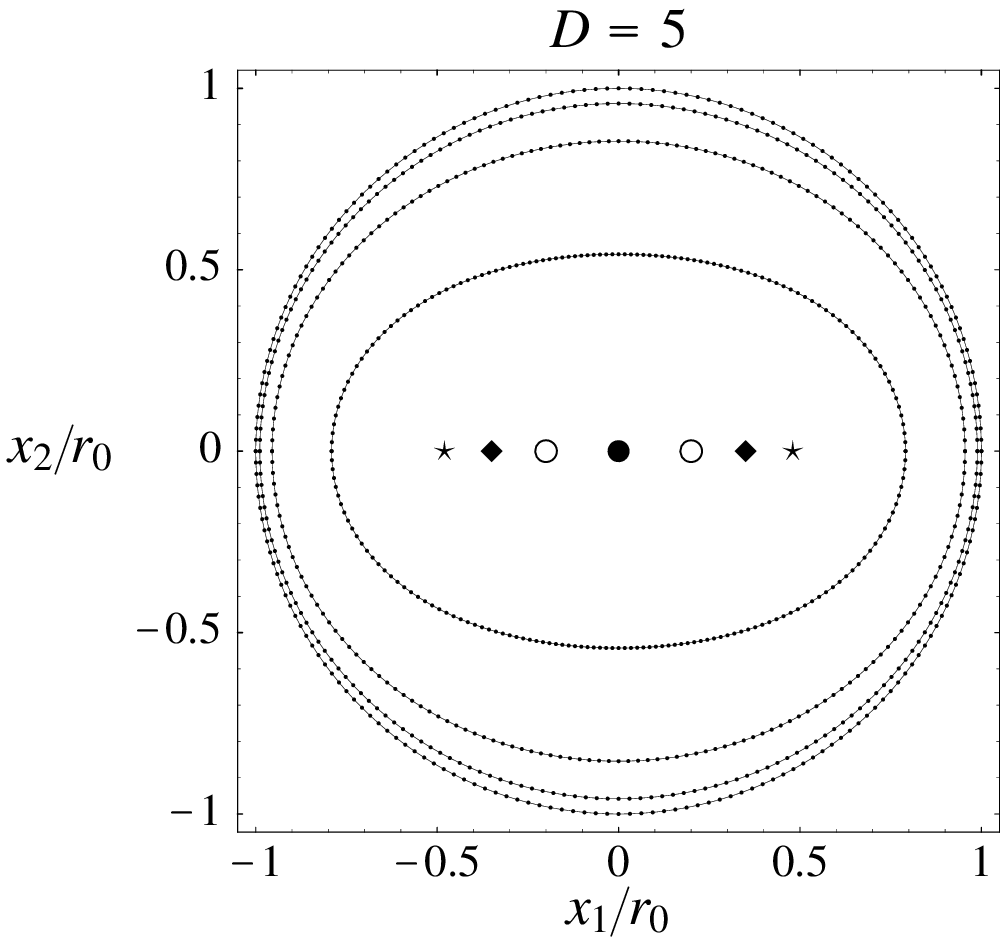}
\includegraphics[width=0.4\linewidth]{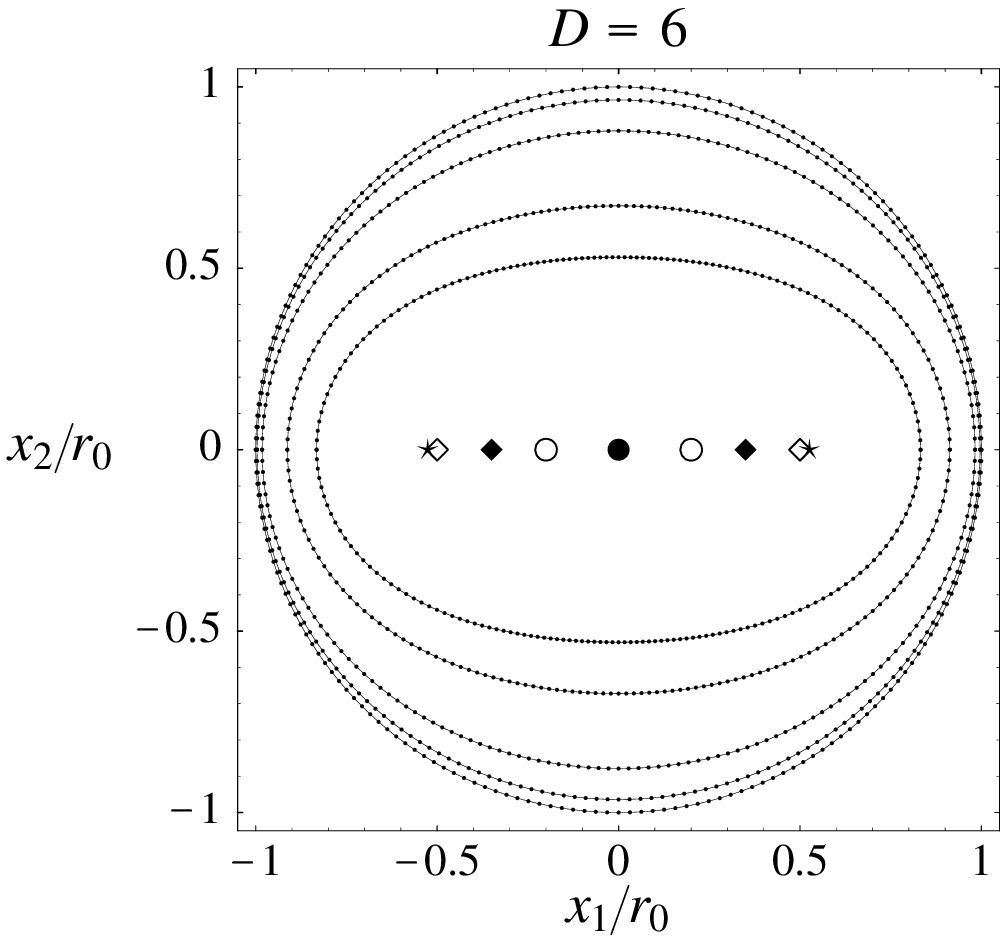}\hspace{10mm}
\includegraphics[width=0.4\linewidth]{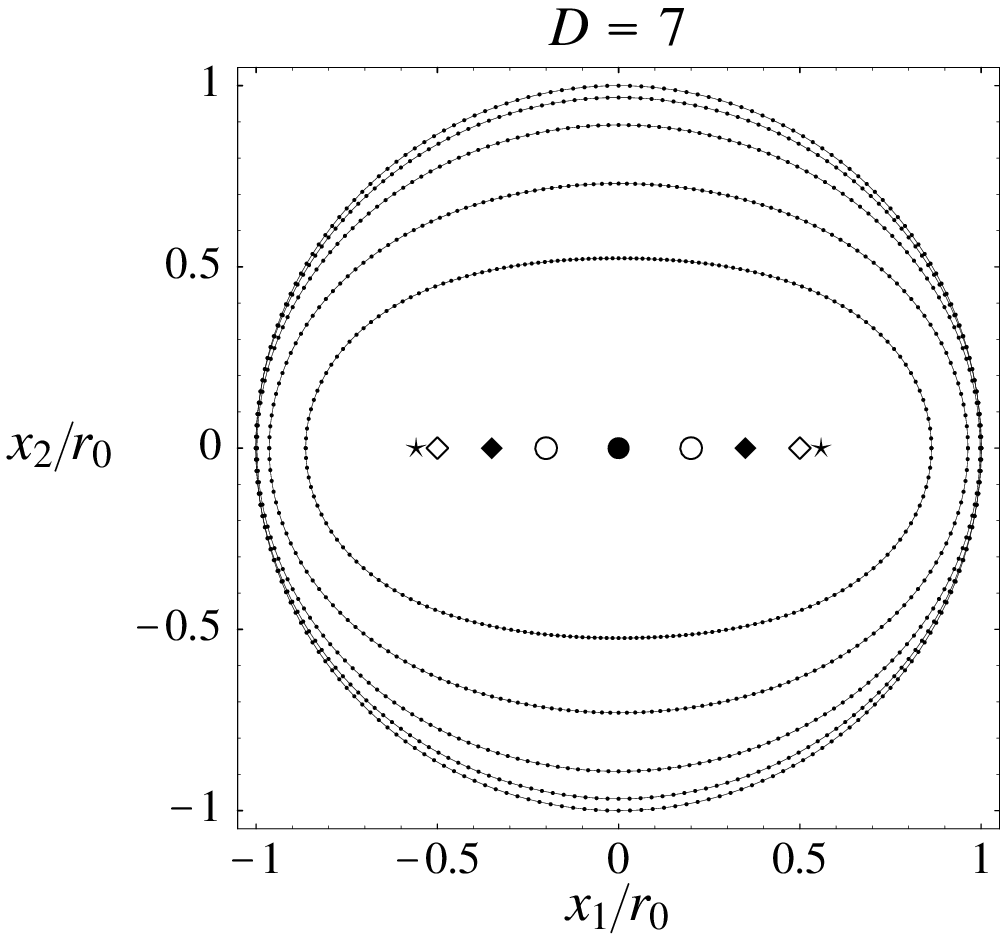}}
\caption{
  The shape of the apparent horizon $\mathcal{C}$ on $(x_1,x_2)$-plane
  in the collision plane $u=v=0$ for $D=4,...,7$.  Incoming particles
  are located on the horizontal line $x_2=0$.  Values of $b/r_0$ are
  $0$($\bullet$), $0.4$($\circ$), $0.7$($\blacklozenge$),
  $1.0$($\lozenge$, shown only in $D=6,7$), and
  $b_{\text{max}}/r_0$($\star$).  As the distance $b$ between two
  particles increases, the radius of $\mathcal{C}$ decreases.  }
\end{figure}
\begin{figure}[ht]
\centering
\includegraphics[width=0.6\linewidth]{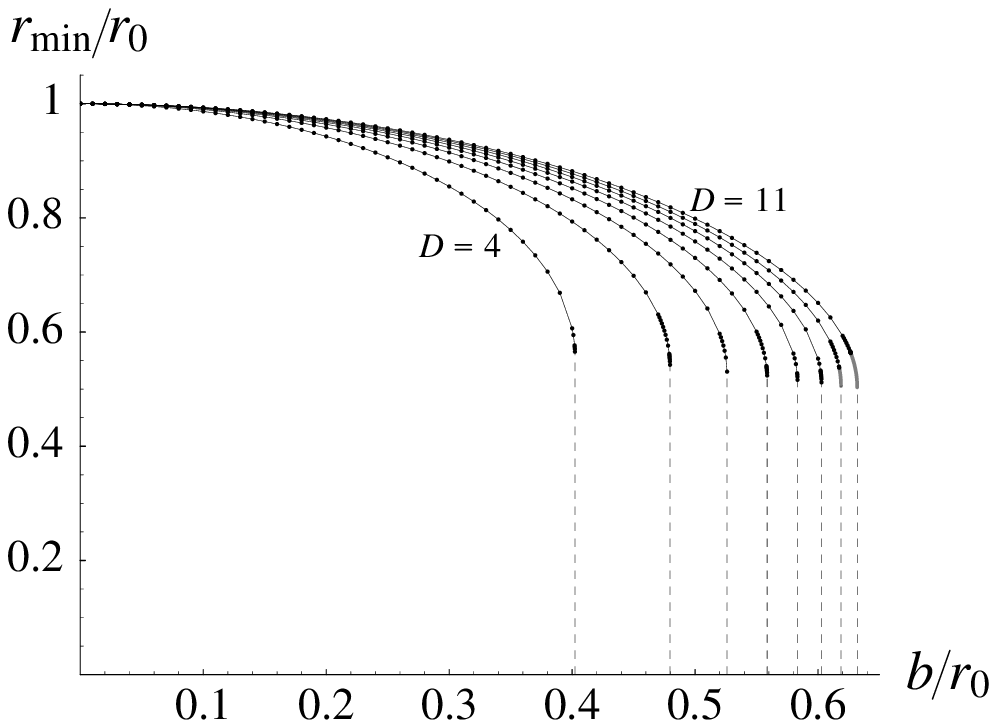}
\caption{
  The relation between the impact parameter $b$ and the minimum radius
  $r_{\text{min}}$ of $\mathcal{C}$ for $D=4,...,11$.  The value of
  $b_{\text{max}}/r_0$ grows as $D$ increases.  }
\end{figure}

Figure 1 shows the shape of $\mathcal{C}$ on $(x_1,x_2)$-plane for
$D=4,..., 7$. For $b=0$, the apparent horizon consists of two 
$(D-2)$-dimensional flat disks whose radius $r_0$ is
\begin{equation}
  r_0=\left(\frac{8\pi\mu G_D}{\Omega_{D-3}}\right)^{1/(D-3)}.
\end{equation}
The radius $g(\theta)$ of $\mathcal{C}$ takes the maximum value
$r_{\text{max}}$ at $\theta=0,\pi$ and takes the minimum value
$r_{\text{min}}$ at $\theta=\pi/2$.  As the impact parameter $b$
increases, the radius of the $(D-3)$-surface $\mathcal{C}$ becomes
smaller.  The shape of $\mathcal{C}$ at $b=b_{\text{max}}$ deviates
from a sphere and the ratio $r_{\text{max}}/r_{\text{min}}$ becomes
large as $D$ increases and more prolate shaped horizon can exist for
larger $D$.

Figure 2 shows the relation between $b$ and $r_{\text{min}}$ for each
$D$. The value of $b_{\text{max}}/r_0$ ranges between $0.8$ and $1.3$
and becomes large as $D$ increases.  Because $r_h(2\mu)\simeq r_0$,
$b_{\text{max}}/r_h(2\mu)$ is also $O(1)$ and increases with $D$.
This behavior of $b_{\text{max}}/r_h(2\mu)$ is different from the case
of the head-on collision~\cite{YN02}, where the ratio $(\text{distance
  of two particles})/r_h(2\mu)$ at the horizon formation decreases with
$D$.  Our numerical results for $b_{\text{max}}/2r_0$ and $F(D)$ are
summarized in Table I.

\begin{table}[tb]
\caption{ The value of $b_{\text{max}}/2r_0$ and $F(D)$ for $D=4,..., 11$.}
\begin{ruledtabular}
\begin{tabular}{c|cccccccc}
  $D$ & 4 & 5 & 6 & 7 & 8 & 9 & 10 & 11 \\
  \hline 
  $b_{\text{max}}/2r_0$ & 0.402 & 0.480 & 0.526 & 0.559 & 0.583 & 0.603
  & 0.619 & 0.632 \\
  $F(D)$ & 0.647 & 1.084 & 1.341 & 1.515 & 1.642 & 1.741 & 1.819 &1.883
\end{tabular}
\end{ruledtabular}
\end{table}

\begin{figure}[t]
\centering
\includegraphics[width=0.6\linewidth]{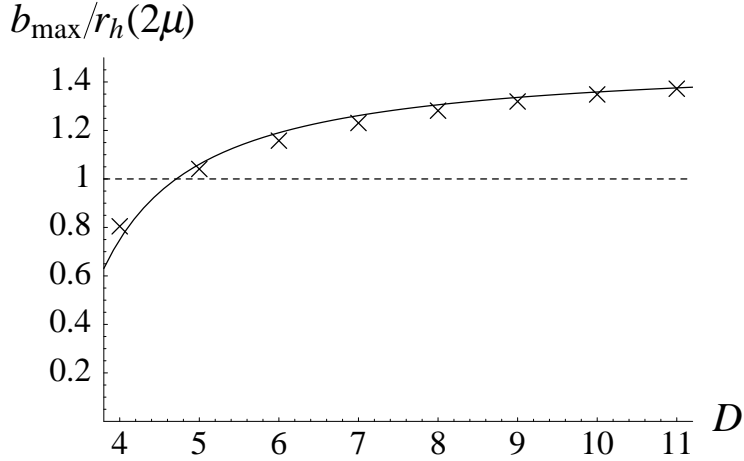}
\caption{
  The value of $b_{\text{max}}$ (crosses) as a function of the
  spacetime dimension $D$.  The dotted line $r_h(2\mu)$ which comes
  from the hoop conjecture. Although the ratio
  $b_{\text{max}}/r_h(2\mu)$ takes the value around unity, the hoop
  conjecture does not explain the increase of
  $b_{\text{max}}/r_h(2\mu)$ with $D$. The solid line
  $b_{\text{max}}/r_h(2\mu)=1.5r_h(\mu)/r_h(2\mu)\sim 2^{-1/(D-3)}$
  gives a good fit of $b_{\text{max}}$.}
\end{figure}

Now we investigate the condition for horizon formation in the two
particle system. Because $b_{\text{max}}/r_h(2\mu)\sim O(1) $, the
hoop conjecture can be applicable to estimate the maximal impact
parameter in high-energy collisions in the higher-dimensional
spacetime. But the hoop conjecture cannot explain the increase of
$b_{\text{max}}/r_h(2\mu)$ with the increase of $D$.  Figure 3 is the
plot of $b_{\text{max}}(D)$.  We found that the relation
$b_{\text{max}}\simeq 1.5r_h(\mu)$ well describes the behavior of
$b_{\text{max}}(D)$ obtained by numerical calculation.  As $r_h(\mu)$
is an effective gravitational radius of each incoming particle, the
horizon formation occurs when there is an overlap of more than one
half of $r_h(\mu)$ between the regions with radius $r_h(\mu)$ around
each particles.  Because $r_h(\mu)$ is proportional to
$\mu^{1/(D-3)}$, the ratio $r_h(\mu)/r_h(2\mu)=2^{-1/(D-3)}$ increases
with $D$. This leads to the increase of $b_{\text{max}}/r_h(2\mu)$
with $D$.

The behavior $b_{\text{max}}/r_h(2\mu)\propto 2^{-1/(D-3)}$ can be
derived from the more fundamental principle, the $(D-3)$-dimensional
volume conjecture. In the present system, the condition \eqref{eq:vc}
can be written as
\begin{equation}
\mathcal{H}_{D}\equiv 
\left[\frac{V_{D-3}}{\Omega_{D-3}r_h^{D-3}(M)}\right]^{1/(D-3)}\lesssim 1.
\label{eq:vdm3}
\end{equation}
This condition reduces to the hoop conjecture \eqref{eq:hc} for $D=4$ case.  
As the characteristic length of the system is $b$ in $x_1$-direction and
$r_h(\mu)$ in other directions, the characteristic $(D-3)$-volume
becomes $V_{D-3}\sim b\,r_h^{D-4}(\mu)$ and $\mathcal{H}_{D}$ becomes
$\sim \left[{br_h^{D-4}(\mu)}/{r_h^{D-3}(2\mu)}\right]^{1/(D-3)}$. By
setting $\mathcal{H}_{D}\sim 1$, the maximal impact parameter is given
by
\begin{equation}
  \label{eq:estimate}
  \frac{b_{\text{max}}}{r_h(2\mu)}
  \sim \left(\frac{r_h(2\mu)}{r_h(\mu)}\right)^{D-4}
  \propto 2^{-1/(D-3)}.
\end{equation}
Although this is rough order estimation, the calculation indicates
that the volume conjecture provides a better condition than the hoop
conjecture.

\begin{figure}[t]
\centering
\includegraphics[width=0.6\linewidth]{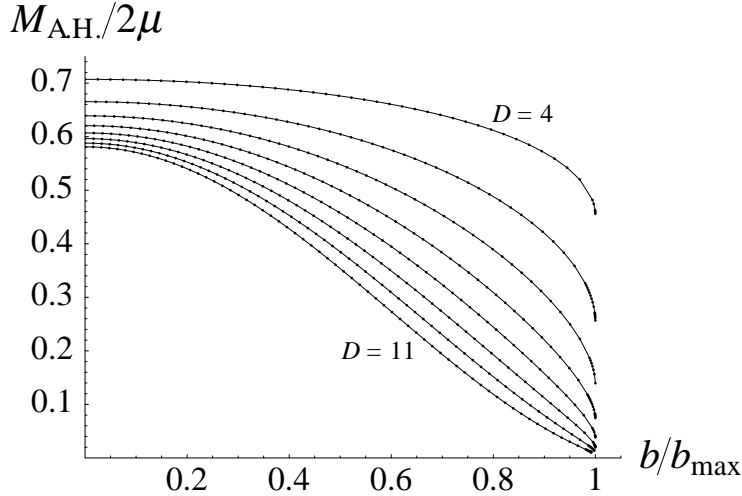}
\caption{
  The relation between the horizon mass $M_{\text{A.H.}}/2\mu$ and the
  impact parameter $b$ for $D=4,..., 11$. $M_{\text{A.H.}}/2\mu$
  becomes small as $D$ increases.  The black hole in the higher
  dimensional spacetime can trap small amount of the energy.  }
\end{figure}

To test the validity of the two conjectures further, we consider how
much energy is trapped by the black hole. This is related to the
definition of the mass in the system.  In our previous
paper~\cite{YNT01}, we discussed the existence of the gravitational
wave energy that does not contribute to the horizon formation in the
system with motion and found that the hoop conjecture with Hawking's
quasilocal mass provides a good condition for the horizon formation
in four-dimensional gravity.  In the present system, we calculate the
quantity
\begin{equation}
  M_{\text{A.H.}}\equiv\frac{(D-2)\Omega_{D-2}}{16\pi G_D}
  \left(\frac{A_{D-2}}{\Omega_{D-2}}\right)^{{(D-3)}/{(D-2)}},
\label{eq:area}
\end{equation}
where $A_{D-2}$ is $(D-2)$-dimensional area of the apparent horizon.
By the area theorem, $M_{\text{A.H.}}$ provides the lower bound of the
final mass of the black hole and coincides Hawking's quasilocal mass
on the horizon.  Therefore $M_{\text{A.H.}}$ can become an indicator
of the energy trapped by the horizon.  Figure 4 shows the behavior of
$M_{\text{A.H.}}/2\mu$ as a function of $b$ for each $D$. We find that
the value of $M_{\text{A.H.}}/2\mu$ at $b=b_{\text{max}}$ decreases as
$D$ increases.  In the higher-dimensional spacetime, the amount of
``junk'' energy increases because the gravitational field distributes
in the space of the extra-dimensions and only a small portion of the
total energy of the system can contribute to the horizon formation.
This junk energy will be radiated away rapidly after the formation of
the black hole.  This strongly suggests that we should use the
quasilocal mass to check the hoop and the volume conjectures.

To test the validity of the two conjectures with the quasilocal mass,
we must calculate $M(\mathrm{S})$, $C(\mathrm{S})$ and
$V_{D-3}(\mathrm{S})$ for all surfaces S, and then take the minimum
value of $C(\mathrm{S})/2\pi r_h(M(\mathrm{S}))$ and
$V_{D-3}(\mathrm{S})/\Omega_{D-3}r_h^{D-3}(M(\mathrm{S}))$.  In the
present system, calculation of the quasilocal mass is difficult
because we treat the black hole on two connected null hyperplanes,
while ordinary quasilocal mass is defined in spacelike hypersurface.
Hence we calculate the value of $H_D^{\text{A.H.}}\equiv
C(\mathrm{S}_{\text{A.H.}})/2\pi r_h(M_{\text{A.H.}})$ and
$\mathcal{H}_{D}^{\text{A.H.}}\equiv
\left[V_{D-3}(\mathrm{S}_{\text{A.H.}})/
  \Omega_{D-3}r_h^{D-3}(M_{\text{A.H.}})\right]^{1/(D-3)}$ on the
apparent horizon $\mathrm{S}_{\text{A.H.}}$.
\begin{figure}[t]
\centering
\includegraphics[width=0.6\linewidth]{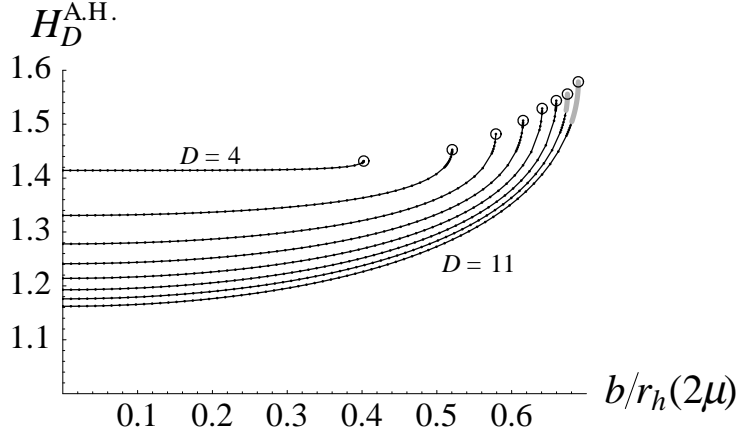}
\caption{
  The value of $H_D^{\text{A.H.}}$ as a function of $b/r_h(2\mu)$ for
  $D=4,..., 11$. The circles show the values at $b=b_{\text{max}}$.
  $H_D^{\text{A.H.}}(b_{\text{max}})$ becomes large as $D$ increases.
}
\end{figure}
\begin{figure}[t]
\centering
\includegraphics[width=0.6\linewidth]{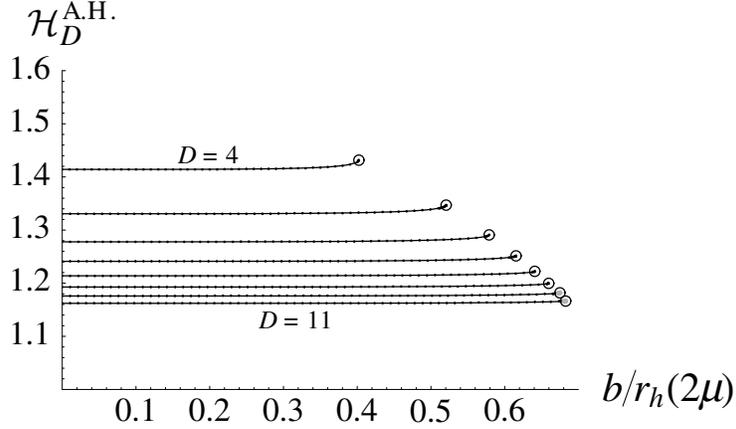}
\caption{
  The value of $\mathcal{H}_D^{\text{A.H.}}$ as a function of
  $b/r_h(2\mu)$ for $D=4,..., 11$. The circles show the values at
  $b=b_{\text{max}}$.  $\mathcal{H}_D^{\text{A.H.}}(b_{\text{max}})$
  becomes small as $D$ increases.  }
\end{figure}
In calculating $H_D^{\text{A.H.}}$, we must specify the hoop for the
surface $\text{S}_{\text{A.H.}}$.  The hoop is defined as the longest closed
geodesic on the surface.  In our system, the hoop is the length of the
curve $\mathcal{C}$ on $(x_1, x_2)$-plane shown in Figure 1.  Figure 5
shows the numerical results of $H_D^{\text{A.H.}}(b)$ for each $D$.
$H_D^{\text{A.H.}}(b)$ monotonically increases with $b$ for all $D$
and it can be a parameter to judge the existence of the horizon.  The
value at horizon formation $H_D^{\text{A.H.}}(b_{\text{max}})$ ranges
between $1.4$ and $1.6$ and increases with $D$. The deviation of the
value $H_D$ from unity is due to the definition of the mass and we
expect that this value approaches close to unity if we use the
appropriate mass of the system and the hoop conjecture holds.  The
increase of $H_D^{\text{A.H.}}(b_{\text{max}})$ with $D$ reflects the
fact that more prolate shaped horizon can form for larger $D$ and
$b_{\text{max}}/r_h(2\mu)$ increases with $D$.

As for $(D-3)$-volume, general definition has not existed yet.  Ida
and Nakao\cite{IN02} selected some $(D-3)$-surfaces by considering
symmetries of the horizon and took the maximum value of their
$(D-3)$-volume.  Using the similar method, we adopt the $(D-3)$-volume
of the $(D-3)$-surface $\mathcal{C}$.  Figure 6 shows the numerical
results of $\mathcal{H}_{D}^{\text{A.H.}}(b)$.
$\mathcal{H}_{D}^{\text{A.H.}}(b)$ is almost constant and slightly
increases with $b$:
\begin{equation}
  \mathcal{H}_{D}^{\text{A.H.}}(b)
  \simeq \mathcal{H}_{D}^{\text{A.H.}}(0)
  =\left[\frac{(D-2)\Omega_{D-2}}{2\Omega_{D-3}}\right]^{1/(D-2)}.
\label{eq:dm3vAH}
\end{equation} 
Thus the isperimetric inequality
$\mathcal{H}_{D}^{\text{A.H.}}(b)\lesssim 1$ holds on the horizon in
this system.  The factor in Eq.\eqref{eq:dm3vAH} contradicts the
constancy of $\mathcal{H}_{D}^{\text{A.H.}}(b_{\text{max}})$ with
respect to $D$ that we have expected from \eqref{eq:estimate}. It may
be absorbed to the definition of the volume factor, but we do not have
the rigorous definition of the volume and we cannot explain the
meaning of this factor at this stage. By ignoring this factor, the
behavior of $\mathcal{H}_{D}^{\text{A.H.}}(b)$ for each $D$ is similar
to $H_D^{\text{A.H.}}(b)$ for $D=4$.  This indicates that
$\mathcal{H}_D$ with the quasilocal mass provides the value that is
close to unity at the horizon formation and the volume conjecture
holds.

\section{Summary and discussion}

We have investigated the black hole formation in the grazing collision
of high-energy particles numerically.  The black hole is produced when
the impact parameter $b$ satisfies the condition $b\lesssim 1.5
r_h(\mu)$.  This condition can be derived by using the volume
conjecture and suggests that the volume conjecture provides a better
condition for horizon formation than the hoop conjecture in the
higher-dimensional gravity.

We evaluated the value of $H_D$ and $\mathcal{H}_{D}$ on the apparent
horizon and both inequalities $H_D\lesssim 1$ and
$\mathcal{H}_D\lesssim 1$ give fairly good conditions for the horizon
formation.  This result is in contrast to the horizon formation by the
spindle matter distribution \cite{IN02}; In this case, the
gravitational field in the transverse direction of spindle matter is
$(D-1)$-dimensional and the black hole forms even if the hoop $C$ is
far greater than $2\pi r_h(M)$ for $D\geq 5$ and $H_D\lesssim 1$ does
not give the condition for the horizon formation.  But in the system
with two point particles, the gravitational field in the middle of two
particles is weak if the distance of two particles are larger than
$r_h(M)$ and becomes sufficiently strong to form the horizon if the
distance of two particles equals to $r_h(M)$ for any $D$.  This leads
to the condition of the black hole formation $H_D\sim 1$.  We
therefore expect that the hoop conjecture in the higher-dimensional
spacetime holds for the system that consists of two point particles.

The behavior of $\mathcal{H}_D^{\text{A.H.}}(b_{\text{max}})$
indicates that $\mathcal{H}_D\lesssim 1$ with the quasilocal mass
provides the better condition for the horizon formation compared to
the hoop conjecture. But what we have confirmed is the necessary
condition for the horizon formation; if an apparent horizon exists,
there is a surface that satisfies $\mathcal{H}_D\lesssim 1$.  There
remains the possibility that even if the apparent horizon does not
exist, there is a surface that satisfies the $\mathcal{H}_D\lesssim
1$.  Hence further investigation is required to confirm whether
$(D-3)$-volume conjecture provides a sufficient condition for the
horizon formation.  This is our remaining problem.


\end{document}